\documentclass[12pt]{iopart}

\usepackage{graphicx}

\begin{document}

\title[Fermi level density of states modulation without charge
  transfer]{Fermi level density of states modulation without charge
  transfer in nickelate superlattices}

\author{Myung Joon Han $ˆ{1,2,3}$ and Michel van Veenendaal $ˆ{2,3}$}
\address{$ˆ1$ Department of Physics, Korea Advanced Institute of
  Science and Technology, Daejeon 305-701, Korea }
\address{$ˆ2$ Advanced Photon Source, Argonne National Laboratory,
  Argonne, Illinois 60439, USA}
\address{$ˆ3$ Department of Physics, Northern Illinois University, De
  Kalb, Illinois 60115, USA} 
\ead{mj.han@kaist.ac.kr}


\date{\today }

\begin{abstract}
By using first-principles density
functional theory calculations for (LaNiO$_3$)$_m$/(SrTiO$_3$)$_n$
superlattices, we report a systematic way of electronic response
to the interface geometry. It is found that Fermi level
density of states of metallic nickelate layers is significantly
reduced without charge transfer in the vicinity of interface to the
insulating SrTiO$_3$. 
This type of electronic state redistribution is clearly distinctive
from other interface phenomena such as charge and orbital reconstruction.
 Our result sheds new light towards understanding
the nickelates and other transition-metal oxide heterostructures.
\end{abstract}


\maketitle

\section{Introduction}
Recent advances in the layer-by-layer growth technique of
transition-metal oxide (TMO) heterostructures have created
considerable research interest \cite{MRS,Hwang_review}. In these artificially
structured interfaces, many exotic material characteristics have been
reported that are strikingly different from the bulk properties and
 the characteristics of a typical semiconducting interface
(non-TMO interface) \cite{MRS,Hwang_review}.  For example, early studies
of Ti-based superlattices such as LaTiO$_3$/SrTiO$_3$ (LTO/STO) and
SrTiO$_3$/LaAlO$_3$ (STO/LAO) have revealed that heterostructuring
induces unique electronic processes in TMOs that can dramatically
change the macroscopic material properties
\cite{Ohtomo-1,Okamoto,Ohtomo-2,Nakagawa}. In these cases, it is
widely believed that the valence charge reconstruction and the polar
discontinuity can drive the interfaces to be metallic even if the two
mother materials are good insulators. The orbital degrees of freedom
are also found to be reconstructed at the interface between manganite
and cuprate \cite{OrbReconst}. Other examples of emergent material
properties caused by the heterostructure geometry include
superconductivity \cite{Reyren} and magnetism
\cite{mag-2007,mag-LAOSTO-1,mag-LAOSTO-2}.  These new findings raise
the important question how the heterostructuring of TMO materials
affects the density of states (DOS) close to the Fermi level and the
related low-energy properties.

In this study, we report another type of electronic response to the
interface geometry, that is, Fermi level DOS modulation caused by the
state redistribution without valence change.  Although this
phenomenon exhibits similarities to the charge and orbital
reconstruction, it displays distinctive characteristics.  We found
that the Fermi level DOS strongly modulates as a function of the
location of nickel atoms relative to the interface.  Our
first-principles calculations of LaNiO$_3$ (LNO)/STO clearly
demonstrate that this novel electronic modulation around the Fermi
level is intrinsic to the structure itself, {\it i.e.,} the
heterostructuring of nickelates, and not originating due to the other
effects such as the charge transfer, oxidation, and valence change.
Further analysis indicates that this kind of behavior can be a common
feature of the nickelate system sandwiched by {\it any} wide-gap
material.  Our result sheds new light on the understanding of TMO
interface phenomena.

It should be noted that (LNO)$_m$/(STO)$_n$ is distinctive from the
widely-studied systems such as LNO/LAO and LNO film
\cite{Son-APL,Hansmann-PRB,MJHan-opol,MJHan-slab,Jian-PRB-R,MJHan-DMFT,MJHan-LDAU,Sakai-PRB}
due to the possibility that Ti can have active $d$ electrons around
the Fermi level. The charge transfer can in principle take place
between Ti and Ni. As an example, it would be instructive to compare
LNO/STO to LNO/LTO superlattice in which the Ti can clearly have an
electron in its $d$ orbitals, and there may be the electron transfer
from Ti to Ni, leading to the configuration of Ti as $d^{0+\delta}$
(or $d^{1-\delta}$ depending on the amount of charge transfer) and Ni
as $d^{8-\delta}$ (or $d^{7+\delta}$) \cite{Hanghui}. This kind of
charge transfer is known to play an important role in determining the
magnetic property of LNO/LaMnO$_3$ (LMO)
\cite{ex-bias,Hoffman,Dong,ATLee}. However, in the case of LNO/STO, it
is not clear if such an electron transfer would take place especially
for the case of $m$=$n$=1. Note that (LNO)$_1$/(STO)$_1$ can also be
identified as (SrNiO$_3$)$_1$/(LTO)$_1$, and this alternative
specification of the system implies that Ti has $d^{1}$ (LTO-like)
configuration instead of $d^0$ (STO-like). As we will show in the
below, this possibility is not realized and Ti remains as $d^0$ even
in the $m$=$n$=1 case, and Ni as $d^7$ (we drop off the indication of
ligand hole for simplicity). From this interesting finding, important
questions arise: If there is no charge transfer between the transition
metal (TM) ions, what happens at the interface?  In other words, under
the condition that no charge transfer is allowed, what kind of
response can be made by the metallic material, LNO, in the vicinity of
interface to the insulating STO? The interface Ni would exhibit the
same characteristics with the inner layer Ni?  If not, what kind of
possibility does Ni have in this superlattice geometry?  This is a
well defined open question that has never been addressed clearly
before.

Our calculations show that the Ni-$d$ states are actually adjusted
at the interface in a systematic way that its DOS is redistributed
while keeping the same total number of electrons. That is, the
interface Ni, closer to the insulating STO, becomes more
insulator-like in the sense that its Fermi level DOS gets reduced
while the inner layer Ni more metal-like in the sense that the more
DOS at the Fermi energy.

\begin{figure*}[t]
\begin{center}
\includegraphics[width=6cm,angle=270]{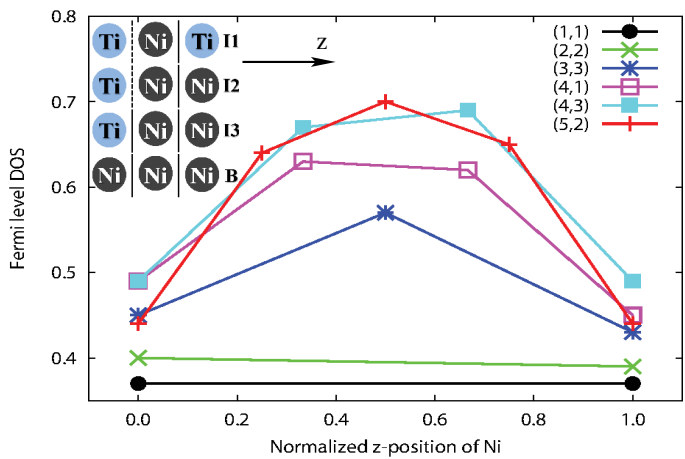}
\caption{ The calculated Fermi level DOS for various ($m,n$) combinations
  of superlattices.  The $z$-axis positions of Ni ions are
  renormalized so that the first NiO$_2$ layer is located at $z=0$ and
  the last is located at $z=1$.  (Inset) Four different types of Ni
  sites existing in the (LNO)$_m$/(STO)$_n$ superlattice. Dark gray
  and light blue circles represent Ni and Ti, respectively. Solid and
  dotted vertical bars in between transition metals represent SrO and
  LaO layers, respectively. Nickel of type I1 is located in between
  the TiO$_2$-SrO and LaO-TiO$_2$ layers, type I2 in between
  TiO$_2$-SrO and LaO-NiO$_2$, and type I3 in between NiO$_2$-LaO and
  LaO-TiO$_2$. Nickel of type B is in the bulk-like arrangement
  located in between NiO$_2$-LaO and LaO-NiO$_2$.
   \label{combined}}
\end{center}
\end{figure*}

\section{Computation Details}
For band structure calculations, we employed the norm-conserving
pseudopotential with a partial core correction, and a linear
combination of localized pseudo-atomic orbitals (LCPAO) as a basis set
\cite{OpenMX}.  We adopted the local density approximation (LDA) for
the exchange-correlation energy functional as parametrized by Perdew
and Zunger \cite{CA}, and used an energy cutoff of 400 Ry and k-grid
of $12\times 12\times 6$ per unit volume. It is noted that bulk LNO is
a paramagnetic metal down to low temperatures and that there is no
report yet on the magnetic order in LNO/STO even if a theoretical
calculation discussed the other possibilities \cite{HSKim}.
Importantly, a recent standing-wave x-ray photoemission spectroscopy
(SWXPS) measurement also assumed the paramagnetic phase
\cite{Kaiser-PRL}, which will serve as a main experimental reference
for our calculation.  Therefore, a conventional LDA can be the best
choice not only because it has been widely used in the previous
studies to describe the paramagnetic LNO, but also because the LDA+U
type of calculation has to assume a long-range ordered magnetic phase.

We used the Mulliken method for the charge analysis in which the
Kohn-Sham states are projected onto our LCPAO basis orbitals. The
number of Ni-$e_g$ electrons obtained in this study is in the range
comparable to the case of LNO/LXO (X: B, Al, Ga, In) superlattice
systems as reported in Ref.~21.  The geometry relaxation has been
performed with the force criterion of $10^{-3}$ Hartree/Bohr. During
the relaxation process, we assume that the in-plane lattice constant
does not change due to its pinning with the substrate (we assumed the
STO lattice parameter for the substrate).

\section{Result}

\subsection{Classification of Ni types and modulation of Fermi level DOS}

In LNO/STO, four different Ni sites exist considering their local
environment.  Nickel can be located either in between two
La$^{3+}$O$^{2-}$ layers or in between the La$^{3+}$O$^{2-}$ and
Sr$^{2+}$O$^{2-}$ layers. Further, its neighboring TM can be either
titanium or nickel (see the inset of Fig.~\ref{combined}).  If a
nickel ion is located in between two La$^{3+}$O$^{2-}$ layers and its
two neighboring TM sites are both Ni, it is bulk-like. At the
interface, there are three different local structures denoted as I1,
I2, and I3 in the inset of Fig.~\ref{combined}. As the electronic
structure of Ni can be affected by both of these two factors (the
neighboring TM ions and the ionic potentials caused by $A$-site
cations), it is important to understand these effects on the Ni
electronic structures. We will discuss this point below in further
detail.

Recent depth-resolved SWXPS studies \cite{Kaiser-PRL} show that the
near-Fermi-level nickel states in (LNO)$_4$/(STO)$_3$ are
significantly suppressed at the interface. A detailed analysis of the
angular behavior of the SWXPS spectra indicates the electronic states
reduction at the two outer LNO layers adjacent to the STO layers (type
I2 and I3 nickel in our definition), but not in the inner layers
(bulk-like nickel). It was speculated in Ref.~28 that the variation in
Fermi level DOS could be due to the different oxidation of Ni ions;
however, no further experimental evidence or explanation was provided.

\subsection{DOS redistribution without charge transfer}
We note that a clear understanding of this DOS modulation is of
significant importance. If there is a charge transfer between TM ions,
it is quite natural to expect that the amount of DOS changes at the
Fermi level. Interestingly, however, our calculation shows that such a
charge transfer and oxidation process are not responsible for this DOS
modulation. We performed calculations for
(LNO)$_m$/(STO)$_n$ with $(m,n)=(1,1), (2,2), (3,3), (4,1), (4,3)$,
and $(5,2)$, and found that the number of $e_g$ electrons is
2.36--2.39 regardless of the type of nickel. From the point of view
that all the nickels remain in the same charge status, the observed
DOS modulation is unexpected and hard to be explained by the `electronic
reconstruction' processes that were suggested before.

\begin{figure}[t]
\begin{center}
 \includegraphics[width=6.5cm]{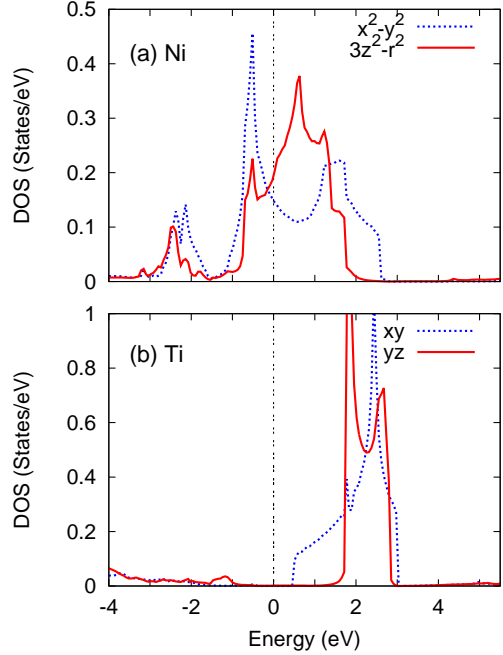}
\caption{ (a) Projected Ni-$e_g$ DOS in the ($1,1$) superlattice. Red
  (solid) and blue (dashed) lines represent the $d_{3z^2-r^2}$ and
  $d_{x^2-y^2}$ states, respectively. (b) Projected Ti-$t_{2g}$ DOS in
  the ($1,1$) superlattice. Blue-dashed and red-solid lines represent
  the $d_{xy}$ and $d_{yz,zx}$ states, respectively.  Vertical dotted
  lines indicate the position of Fermi energy.
    \label{N1T1}}  
\end{center}
\end{figure}

The fact that there is no valence change or further oxidation of Ni
may be most dramatically seen in $m$=$n$=1 case for which the
alternative characterization of the superlattice as
(SNO)$_1$/(LTO)$_1$ is also possible if the charge transfer (from Ni
to Ti) might really happen. Since Ti has $d^1$ configuration in LTO
while $d^0$ in STO, the result can be clearly seen in the Ti
states. The calculated Ti-$d$ DOS in Fig.~\ref{N1T1}(b) shows that all
the Ti-$d$ states are empty.  Therefore one can identify the system as
(LNO)$_1$/(STO)$_1$, and there is no charge transfer between Ni and Ti
even in the $m$=$n$=1 case.

\begin{figure}[t]
\begin{center}
 \includegraphics[width=7cm]{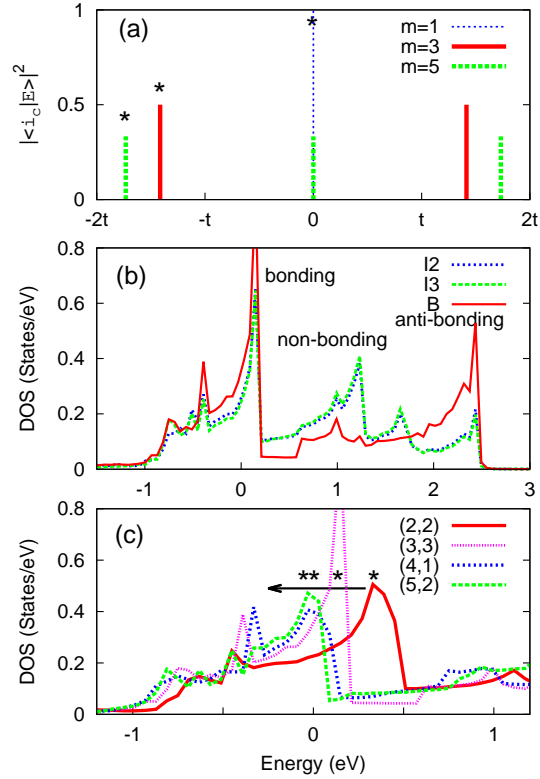}
\caption{ (a) Tight-binding calculation results for the weight of the
  central atom $ |\langle i_c|E \rangle|^2$ in the energy
  eigenfunctions $|E\rangle$. The length of the chain corresponds to
  the thickness of the LNO layer, {\it i.e.,} $m$ in
  (LNO)$_m$/(STO)$_n$. The central position is given by $i_c=m/2$ and
  $(m+1)/2$ for even and odd $m$, respectively. Energy unit is in the
  hopping parameter $t$. (b) Projected $d_{3z^2-r^2}$ DOS for (3,3)
  case. As expected, the bonding, non-bonding, and anti-bonding
  characteristics are most clearly seen in the case of (3,3). Type I2,
  I3, and B states are represented by blue-dotted, green-dashed, and
  red-solid lines, respectively. The Fermi level is set to be
  zero. (c) The projected $d_{3z^2-r^2}$ DOS for bulk-type Ni in the
  (3,3), (4,1), and (5,2) structures. For comparison, type I2 in (2,2)
  is also shown (red). The Fermi level is set to be zero.  The
  asterisk marks indicate the peak positions.
 \label{evolutionDOS_chainModel}}
\end{center}
\end{figure}

\begin{figure}[t]
\begin{center}
 \includegraphics[width=7cm,angle=0]{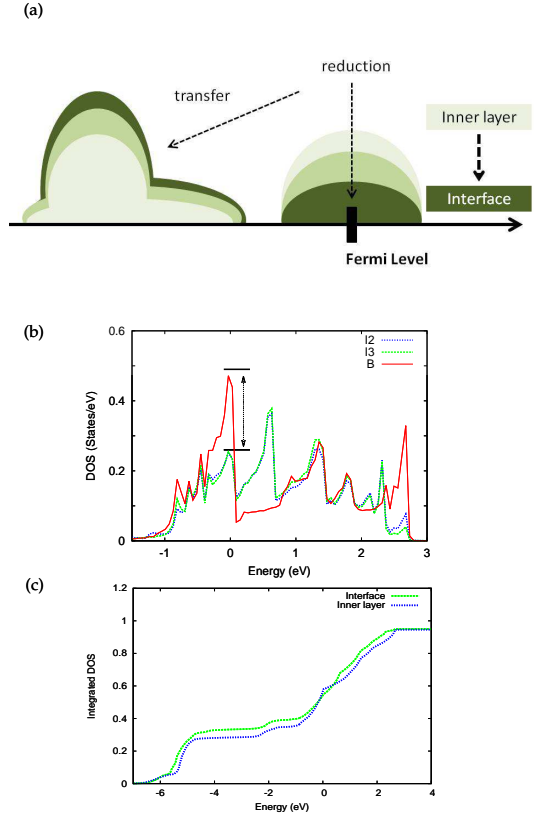}
\caption{ (a) Schematic illustration of DOS reduction at the Fermi
  level as a function of layer position relative to the interface. The
  reduced DOS is transferred to the lower energy part. (b) Projected
  $d_{3z^2-r^2}$ DOS of two interface layer nickels (blue and green)
  and the inner layer nickel (red) in the (5,2) superlattice.  The
  arrow indicates the amount of DOS reduction. (c) The integrated
  $d_{3z^2-r^2}$ DOS for (5,2) case. The interface and inner-most
  layer state are depicted by green and blue color, respectively.
    \label{reduc_DOS2}}
\end{center}
\end{figure}

The effect of $A$-site cation potential asymmetry ({\it i.e.,}
Sr$^{2+}$ versus La$^{3+}$) does not cause any significant change in
the electronic structure.  This effect can be studied by examining the
(3,3) heterostructure, for example, that contains both type I2 and I3
nickel ions (compare the dotted-blue with dashed-green line in
Fig.~\ref{evolutionDOS_chainModel}(b)). Since the two nickel ions only
differ due to the $A$-site cations, the different electronic structure
reflects the effect due to the $A$-site asymmetry.
Figure~\ref{evolutionDOS_chainModel}(b) shows that the DOS change is
small. In addition, for other ($m$,$n$) structures, the projected DOS
for type I2 and I3 nickel is always similar (seen in
Fig.~\ref{reduc_DOS2}(b) for the case of (5,2) or upon comparison of
Fermi level DOS value at $z=0$ and $z=1$ in
Fig.~\ref{combined}). Therefore, the valence reconstruction is not
relevant to this nickelate system presumably due to the strong
covalency.

Now, considering that the nickel valence states are all same
regardless of its position and the cation potential effect, it is
quite surprising that the Fermi level DOS significantly varies.  The
calculated value of this quantity is summarized in Fig.~\ref{combined}
where the DOS values at the Fermi level (projected onto the different
nickel sites) are plotted. It is clear that the interfacial nickel
ions (see Fig.~\ref{combined} where the $z$-position is normalized so
that types I1, I2, and I3 located either at $z=0$ or at $z=1$) always
have notably smaller Fermi level DOS than the bulk-like nickel ($0 < z
< 1$ in Fig.~\ref{combined}).  The difference becomes even more
pronounced as the number of Ni layers increases from $m=$ 3 to $m=$
5. For $m=5$, the Fermi level DOS for the inner most layer Ni ions is
larger by a factor $\sim$2 when compared with that of the interfacial
one. Since there is no charge transfer and no valence change, but a
strong DOS modulation at the Fermi level, DOS should be redistributed
in such a way that the total number of electrons is kept same. The
further analysis shows that DOS is actually redistributed in a
systematic way so that the Fermi level DOS gradually decreases as the
nickelate layer gets closer to the interface. In order to make the
charge valence unchanged, this weight is transferred to the lower
energy part as schematically shown in Fig.~\ref{reduc_DOS2}(a). It is
interesting to note that the metallic LNO layers become more
insulator-like when they come closer to the insulating STO, while
keeping the same valence charge; {\it i.e.}, the smaller number of
states at the Fermi level.  On the other hand, the inner LNO layers
far away from STO, become more metallic in the sense that they have
more DOS at the Fermi level.

Fig.~\ref{reduc_DOS2}(b) shows that the reduction of DOS is quite
significant. As seen in Fig.~\ref{reduc_DOS2}(b), the $d_{3z^2-r^2}$
states of the inner most (B; bulk-like) nickel is much larger than the
two interface nickels (I2 and I3). The arrow indicates the DOS
reduction at the Fermi level. The reduced states are transferred to the
lower energy parts in case of interface nickel sites as clearly shown
in the integrated DOS plot (Fig.~\ref{reduc_DOS2}(c)).

\subsection{Mechanism of redistribution}

To understand the mechanism behind the DOS modulation or
redistribution process, a relatively simple picture can be considered
based on the interactions between the molecule-like quantum states
formed by the heterostructure geometry. As a starting point, let us
consider the $m$=$n$=1 case.  As seen in Fig.~\ref{N1T1}(a), the
bandwidth of $d_{3z^2-r^2}$ state is markedly narrower than that of
the $d_{x^2-y^2}$ state, which is a result of the limited
hybridization along the $z$ direction by the presence of STO
layers. From this characteristics of $d_{3z^2-r^2}$ state, one can
treat it as remnants of the molecular orbital character arising from a
chain of $m$ $d_{3z^2-r^2}$ orbitals along the $z$ direction.  These
molecular-orbital-like features are further confirmed by the DOS shape
of the (3,3) case; Figure~\ref{evolutionDOS_chainModel}(b) clearly
shows the features related to the bonding, antibonding, and nonbonding
states of the $m=3$ chain. It is noted that the DOS of the bulk-type
nickel (solid-red line) is strongest in the bonding and antibonding
features at $\sim$0 and $\sim$2.2 eV.

This point is further supported by our tight-binding analysis, which
captures the essence of the features found in the first-principles
results.  Assuming that the system can approximately be treated as a
linear chain of $m$-molecular orbitals, we constructed a tight-binding
model Hamiltonian corresponding to $N$ atoms, $H=\sum_{i=1}^N t
c_i^{\dagger} c_j$, where $c_i^{\dagger}$ creates an electron at the
$i$th site in the $z$ direction. The weight of the central (bulk-like)
atom $|\langle i_c|E\rangle|^2$ is plotted in
Fig.~\ref{evolutionDOS_chainModel}(a) as a function of energy $E$ (in
the unit of $t$). As the chain length, $m$, increases from $m=1$ to 3
and 5, the states interact with each other and spread out over a wide
energy range. For $m=3$, the two-peak feature of bonding and
anti-bonding is most clearly seen, which is consistent with the
results of our first-principles calculation in
Fig.~\ref{evolutionDOS_chainModel}(b). Another important point in this
analysis is that the bonding-peak position gradually shifts toward the
lower energy region as $m$ increases (marked by the asterisks in
Fig.~\ref{evolutionDOS_chainModel}(a)). Figure
\ref{evolutionDOS_chainModel}(c) shows that this characteristic
behavior of the lower energy part of the $d_{3z^2-r^2}$ states is also
found in the first-principles calculations: As the number of layers
increases, the peak position (marked by asterisks in
Fig.~\ref{evolutionDOS_chainModel}(c)) shifts towards the lower energy
region. It is the electronic origin that leads to a strong modulation
in the number of states at the chemical potential shown in
Fig.~\ref{combined}.

\section{Discussion}

The Fermi level DOS modulation has distinctive features from the other
interface reconstructions. In general, electronic \cite{Okamoto} and
orbital \cite{OrbReconst} reconstructions depend on the relative
on-site energies of interfacial ions leading to a redistribution of
charge between different sites and orbitals. In contrast, Fermi-level
DOS modulations involve a strong redistribution of states within the
same sites and the same orbitals due to the presence of the interface.
Since the low-energy properties are mainly determined by the states
close to the chemical potential, strong position-dependent electronic
properties can be expected and macroscopic properties ({\it e.g.,}
resistivity) should be sensitive to these strong modulations. The
direct observation of this kind of electronic response is difficult as
the charge valence of the interface remains unchanged, and only the
near-Fermi-level DOS should be measured. However, we note that SWXPS
can be used for the verification of our results and that the recent
experiment by Kaiser {\it et al.} for ($m$=4, $n$=3) compares well
with our conclusion \cite{Kaiser-PRL}.

Our analysis can be applied to other superlattice structures with LNO
sandwiched by {\it any} wide-gap insulator in which the hybridization
is strongly blocked only along the $z$ direction. Therefore, the same
type of DOS modulation should be found in the related heterostructures
unless the other types of response occur beforehand. To verify this
point, we calculated the Fermi level DOS for (LNO)$_m$/(LAO)$_n$ with
$(m,n)=(3,3), (4,1), (5,2)$, and found that the same modulations occur
in this different systems.  It is noteworthy that our tight-binding
analysis based on the molecular-orbital chain model is better applied
to the case of LNO/LAO. The DOS modulation can be a universal feature
in nickelate superlattices and possibly also in other interfaces in
between metals and insulators.

Fermi level DOS reduction at the interfaces can provide an interesting
new picture for the nickelate superlattices. We note that the DOS
reduction at the interface is observed for all compositions of
($m\geq$3, $n$), while for $m\leq 2$, only the interface nickel ions
exist and there is no bulk-like one. Interestingly, several
experiments on the nickelate superlattices independently report that
the MIT occurs at $m\approx 3$. The same critical thickness was
commonly observed in LNO/STO \cite{Son}, LNO/LAO \cite{Boris-Science},
LNO/SrMnO$_3$ \cite{May-PRB09}, and LNO/LMO \cite{Hoffman}. It is
therefore temping to relate the metallic and insulating phase to the
bulk-like (B) and interface-like (I1, I2, I3) nickel layers,
respectively, as the bulk-like nickel has more states at the Fermi
energy and it start to appear at $m=3$. Further, a recent experiment
by Boris and co-workers reported that this MIT is accompanied with the
paramagnetic to magnetic transition \cite{Boris-Science}. We note
that, in the Kondo-type screening, the screening strength is governed
by the Fermi level DOS as is clear from the Kondo temperature scale,
$T_K \sim e^{-1/\rho_{0}J_{K}}$, where $\rho_{0}$ refers to the Fermi
level DOS. Therefore, it may be interesting to regard the nickelate
superlattices as a kind of Kondo lattice system where the enhanced
Fermi level DOS is responsible for the metallic conduction and
simultaneously for the screening of local magnetic moments.

In the real experimental situations, there may be other possibilities
for the nickelate layers to be adjusted in the heterostructure
geometry. For example, the distortion of the oxygen octahedra and
charge disproportionation can be important in the thin film LNO
\cite{Chak-PRL-LNOfilm,Jian-PRB-R,HSKim}.  In this system of LNO/STO,
a similar type of ionic displacement can also be realized
\cite{Stemmer-new,HSKim} although this possibility cannot be examined
within our unitcell setup.  Our calculations show that even without
atomic distortion and/or charge transfer, the system has an electronic
way of response to the interface geometry.

Since LDA has a limitation to describe electron correlations, the
effect of correlation in Ni-$3d$ on the DOS modulation can be an issue
for the future study. Here we note the previous dynamical mean-field
study for LNO/LAO-type of superlattice \cite{MJHan-DMFT}, which
indicates that the electronic DOS is not much affected by increasing U
as far as the double-counting energy is properly dealt with. Also
considering the good agreement with SWXPS result, our conclusion is
well justified at least for the paramagnetic and metallic region of
phase space.

\section{Summary}
We report the Fermi level DOS modulation across the LNO layers in
LNO/STO superlattices. This modulation is caused by the
heterostructuring itself with no valence change or oxidation and in
good agreement with a recent SWXPS experiment. It is related to but
clearly distinctive from other interface phenomena such as orbital and
valence reconstruction.  Our analysis demonstrates that this
electronic response to the hetero-interface structure originates from
the novel process of quantum state formation and interactions between
them.  It can be related to MIT in the related systems, providing a
new theoretical aspect to the complex-TMO research.

\section{Acknowledgments}
This work was supported by the U.S. Department of Energy (DOE),
DE-FG02-03ER46097, and NIU's Institute for Nanoscience, Engineering,
and Technology. The work at the Argonne National Laboratory was
supported by the U.S.  DOE, Office of Science, Office of Basic Energy
Sciences under Contract No. DE-AC02-06CH11357. Computational resources
were provided by the National Institute of Supercomputing and
Networking/Korea Institute of Science and Technology Information with
supercomputing resources including technical support (Grant
No. KSC-2013-C2-005).

\end{document}